\documentclass[%
 reprint,
 amsmath,amssymb,
 aps,
]{revtex4-2}

\usepackage{float}
\usepackage{placeins}
\usepackage{color}
\usepackage{graphicx}
\usepackage{dcolumn}
\usepackage{bm}

\begin{document}

\preprint{APS/123-QED}

\title{Mixed Diagnostics for Longitudinal Properties of Electron Bunches in a Free-Electron Laser}

\author{J. Zhu, N. M. Lockmann, M. K. Czwalinna, H. Schlarb}
\affiliation{Deutsches Elektronen-Synchrotron DESY, Notkestr. 85, 22607 Hamburg, Germany }

\date{\today}

\begin{abstract}
Longitudinal properties of electron bunches are critical for the performance of a wide range of scientific facilities. In a free-electron laser, for example, the existing diagnostics only provide very limited longitudinal information of the electron bunch during online tuning and optimization. We leverage the power of artificial intelligence to build a neural network model using experimental data, in order to bring the destructive longitudinal phase space (LPS) diagnostics online virtually and improve the existing current profile online diagnostics which uses a coherent transition radiation (CTR) spectrometer. The model can also serve as a digital twin of the real machine on which algorithms can be tested efficiently and effectively. We demonstrate at the FLASH facility that the encoder-decoder model with more than one decoder can make highly accurate predictions of megapixel LPS images and coherent transition radiation spectra concurrently for electron bunches in a bunch train with broad ranges of LPS shapes and peak currents, which are obtained by scanning all the major control knobs for LPS manipulation. Furthermore, we propose a way to significantly improve the CTR spectrometer online measurement by combining the predicted and measured spectra. Our work showcases how to combine virtual and real diagnostics in order to provide heterogeneous and reliable mixed diagnostics for scientific facilities.
\end{abstract}

\maketitle


\section{INTRODUCTION}

Tuning and optimization of the longitudinal phase space (LPS) of electron bunches are of vital importance for the performance of various scientific facilities such as free-electron lasers (FELs) \cite{Decking2020}, ultrafast electron diffractions (UEDs) \cite{Qi2020}, laser-plasma accelerators (LPAs) \cite{Jalas2021}, plasma wakefield accelerators (PWFAs) \cite{Lindstrom2021}, THz-driven accelerators \cite{Tang2021} and so on. The prerequisite of quickly and accurately manipulating the LPS of an electron bunch is being able to measure and monitor it rapidly and reliably. The LPS of an electron bunch is usually measured directly in the time domain by combining a transverse deflecting structure (TDS) and a dipole spectrometer magnet in a dispersive section \cite{Emma2000, Akre2001}. An LPS image provides rich and important information such as the shape, the current profile and the slice energy spread of the electron bunch. However, this diagnostic method interferes with delivering photons to user experiments and thus cannot be employed online during machine tuning and optimization. The current profile of an electron bunch can also be reconstructed by measuring the coherent transition radiation (CTR) \cite{Stephan2011, Schmidt2020} or the coherent diffraction radiation (CDR) \cite{Lockmann2020} spectrum generated by an electron bunch. Although CTR generation is invasive to an electron bunch, it works as an online diagnostics at facilities such as FLASH, in which a single bunch from a bunch train can be selected for this purpose without interfering with user experiments \cite{Stephan2011}. On the other hand, physics-based beam dynamic simulation plays an important role in understanding the LPS of an electron bunch. However, high-resolution physics-based simulation is extremely time-consuming \cite{Qiang2017} and often does not agree with the measurement very well.

\begin{figure}[h]
	\includegraphics[width=0.5\textwidth]{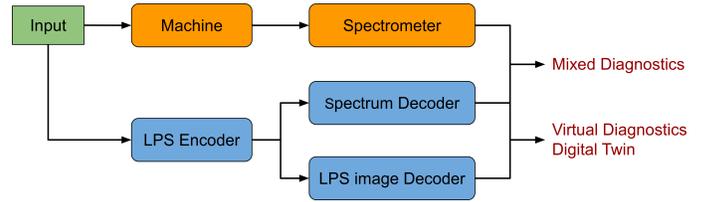}%
	\caption{\label{fig:applications} Applications of an encoder-decoder neural network model with more than one decoder for heterogeneous predictions of electron bunch longitudinal properties.}
\end{figure}
\begin{figure*}[ht]
	\includegraphics[width=1.0\textwidth]{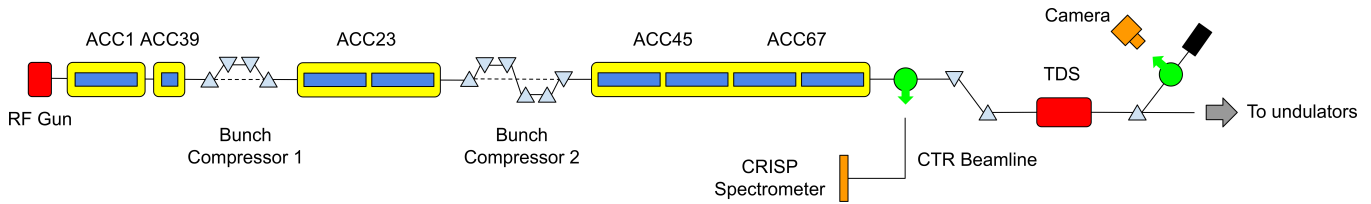}%
	\caption{\label{fig:flash_beamline} Schematic of the FLASH1 beamline. Components are not to scale. ACC1, ACC23, ACC45 and ACC67 are 1.3 GHz cryomodules which boost the electron energy. ACC39 is a 3.9 GHz cryomodule which linearizes the LPS before the first bunch compression stage. The CTR radiation is generated by deflecting one electron bunch onto an off-axis screen using a fast magnetic kicker and measured by the coherent radiation intensity spectrometer (CRISP).}
\end{figure*}

In recent years, machine learning has demonstrated to be able to learn relationships inside a complex system and produce accurate and fast predictions \cite{Anirudh2020, Jumper2021}. Using artificial neural networks as a tool for electron bunch longitudinal property prediction has garnered more and more attention in recent years \cite{Emma2018, Edelen2019, Emma2021, Hanuka2021, Zhu2021}. At the LCLS, early work has demonstrated prediction of LPS images and current profiles at the linac exit using two separated multi-layer perceptrons (MLPs) \cite{Emma2018} and a single input parameter. Later, a spectral virtual diagnostics was proposed to improve the prediction accuracy and robustness by using the CDR spectrum of the electron bunch as input \cite{Hanuka2021}. Simulated CDR spectra were used in this study. Indicated by tests with simulated LPS images for the LCLS-II, a neural network is also expected to predict the microbunching structure in the LPS on a shot-to-shot basis using the spectrum as input with a fixed machine set point \cite{Hanuka2021}. The latest work at the European XFEL injector demonstrated that a deep encoder-decoder neural network can achieve extremely high accuracy in predicting megapixel LPS images using up to three RF phases as input. The current profile, energy spectrum and slice energy spread extracted from the predicted LPS image all show very good agreement with the measurement \cite{Zhu2021}. In addition, an innovative method was demonstrated to efficiently build large models with multiple distinctive working points \cite{Zhu2021}.

%

%
%
%

In this paper, we experimentally demonstrate training an encoder-decoder neural network model with more than one decoder to predict the LPS image and the CTR spectrum of the electron bunches in a single bunch train concurrently at the end of the FLASH linac. Building a model using the data-driven approach heavily relies on the availability of the data. However, the main focus of a user facility is to deliver electron or photon beams to user experiments and thus cannot allocate a large amount of time exclusively for routine data collection. We demonstrate that only a reasonable amount of data is required to train a performant model around a user working point, with all major control knobs for LPS manipulation included. More importantly, we propose a method which can significantly improve the existing online current profile measurement using a CTR spectrometer by combining predicted and measured CTR spectra. Analogous to mixed reality, we call diagnostics which provide a blend of predicted and measured signals mixed diagnostics. This approach offers the possibility to provide heterogeneous and reliable LPS information in real-time for electron bunches with a broad range of parameters.

The applications of the above encoder-decoder model are summarized in Fig.~\ref{fig:applications}. The model can serve as not only virtual and mixed diagnostics for online machine tuning and optimization, but also a digital twin of the actual machine on which machine tuning and optimization algorithms \cite{Scheinker2018, Leemann2019, Duris2020, Bruchon2020, Kain2020, Scheinker2021} can be tested efficiently and effectively before applying them in the real world. These virtual experiments can even be performed in advance to find an optimal or near-optimal setup for the real machine. One of the major advantages of testing with neural network models is that it is orders of magnitude faster than real experiments or physics based simulations because the inference time is typically on the order of milliseconds. Compared with neural network models trained with simulated data \cite{Edelen2020, Hanuka2021-2, Roussel2021}, models trained with experimental data provide better testing environments by generating predictions which are almost identical to the real-world signals.

\begin{figure*}[ht]
	\includegraphics[width=1.0\textwidth]{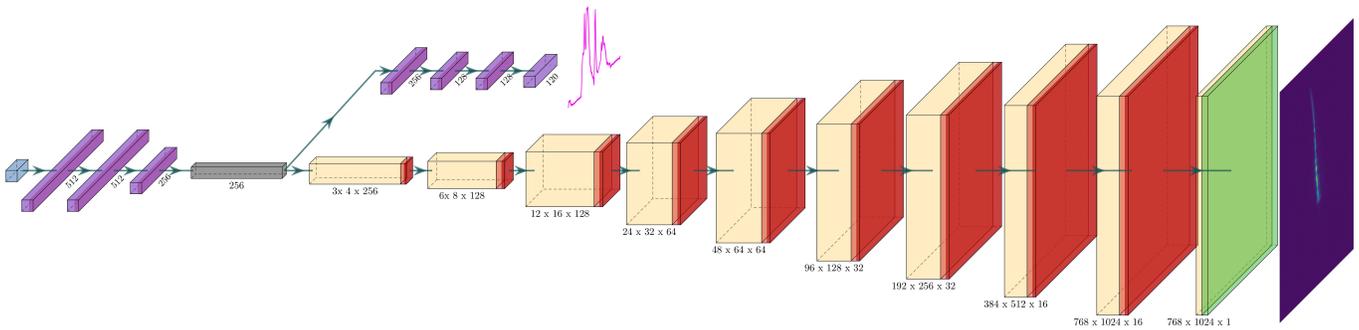}
	\caption{\label{fig:nn} Diagram of the encoder-decoder neural network. The leftmost blue box represents the input layer. It is followed by three fully-connected layers (encoder) in purple with each layer activated by the Leaky ReLU (Rectified Linear Unit) function. The latent space is depicted in grey. The latent space leads to two decoder branches. One consists of several fully-connected layers with each layer activated by the Leaky ReLU function. The other consists of  ten transposed convolutional layers in yellow. Each transposed convolutional layer is followed by a batch normalization layer and activated by the leaky ReLU function except the last one, which is activated by the sigmoid function depicted in green. The kernel sizes of the first and second transposed convolutional layers are 3 x 4 and 3 x 3, respectively, and the kernel sizes of the other eight transposed convolutional layers are all 5 x 5. The total number of trainable parameters is 2,095,145.}
\end{figure*}

\section{Experimental setup}

FLASH is a soft x-ray free-electron laser (FEL) user facility which is capable of delivering MHz pulse trains to two user experiments, FLASH1 and FLASH2, in parallel with individually selected photon beam characteristics \cite{Ackermann2007, Faatz2016}. The layout of the FLASH1 beamline is shown in Fig.~\ref{fig:flash_beamline}. The longitudinal properties of two electron bunches in a bunch train can be measured concurrently by streaking one of the electron bunches using the S-band TDS \cite{ohrs2009, Behrens2012} and picking up another one for the CRISP spectrometer using a fast kicker magnet \cite{Stephan2011}. It should be emphasized that the two diagnostics cannot measure the same bunch in a bunch train simultaneously. In this study, the electron bunch picked up by the CRISP spectrometer travels immediately after the one streaked by the TDS. A systematic comparison between these two diagnostics was conducted recently, which shows excellent agreement on the current profile down to the 10 fs level \cite{Schmidt2020}. However,  due to the non-optimized optics required by the parallel user experiment at the FLASH2 and a cap on the TDS power, the rms time resolution of the measurement using the TDS is larger than 70 fs in our experiments. Moreover, the CRISP spectrometer has two sets of remotely interchangeable grating sets, which cover different frequency ranges. The low-frequency grating set covers the range from 0.7 to 6.6 THz, and the high-frequency grating set covers the range from 6.9 to 58.8 THz. In order to have an accurate reconstruction of the current profile, the full spectra range from 0.7 to 58.8 THz is required. For a meticulous characterization of the current profiles at a few set points, this can be achieved by recording data using the two grating sets consecutively \cite{Schmidt2020}. However, this is not practical for a large number of machine set points. A detailed discussion about how to address this issue when building a neural network model is presented in Sec \ref{mixed_diagnostics}.

Up to 6 control knobs, including the phases and amplitudes of ACC1, ACC39 and ACC23, are scanned during data collection around a user working point because of the following considerations. During machine operation, the RF phases and amplitudes of ACC1, ACC39 and ACC23 are commonly used to adjust the LPS of the electron bunch in order to optimize the FEL performance. The LPS is very sensitive to the phase and amplitude changes of these three RF stations because they affect the LPS of an electron bunch upstream the bunch compressors. ACC45 and ACC67 are operated on-crest downstream the 2nd bunch compressor and thus the LPS is very insensitive to the phase changes of these two RF stations. The amplitudes of ACC45 and ACC67 can be used to slightly tweak the beam energy which have nevertheless negligible impact on the LPS shape.

The scanned parameters are sampled uniformly within predefined ranges, which are mainly restricted by the OTR screen size for the TDS. For each combination, the new values are written into the control system via the Python interface of DOOCS \cite{doocs} and the data readout has a delay of 0.5 second. Since it takes time to collect data which belong to the same bunch train but are sent out from different sources, the actual data collection speed is between 1 and 2 Hz.

\section{Methods}

\subsection{\label{data_wrangling} Data wrangling}

The data quality is essential to the performance of a neural network model. First of all, the data quality in this study is controlled during data taking. A single data point is recorded for each randomly selected machine set point. It prevents data with the same machine set point from appearing in both the train and test data, which leads to overfitting, as much as possible. Secondly, the data are further normalized and cleaned. The original size of the 12-bit camera image is 1360 x 1024 pixels. After background subtraction, the pixels are normalized by 4095. All the pixel values below 0.01 are set to 0 in order to remove negative pixel values and suppress background noise. Although the model is expected to learn and predict the position of the LPS on the screen \cite{Zhu2021}, the horizontal position of the beam (along the streaking direction) depends on a lot of machine parameters, including the phase and amplitude of the TDS. We noticed that the beam sometimes moved significantly horizontally even when the machine set point remained unchanged. As a result, it is not possible to train a performant model even when only a single control knob (e.g. the phase of ACC23) is scanned. Since the horizontal position of the beam only provides the timing information, which can actually be measured using the bunch arrival monitors \cite{Viti2017} installed at different locations of the machine, all the LPS images are cropped to 768 x 1024 and centered horizontally. Moreover, an image will be removed from the dataset if the electron bunch is completely or partially off the screen, or if the electron bunch is very close to the left or right edge.


\subsection{Modeling}

The detailed structure of the encoder-decoder neural network is shown in Fig.~\ref{fig:nn}. A MLP is used to build latent features from the input parameters. Two different decoders then translate the latent features to the LPS image and CTR spectrum (the raw signal measured by the CRISP spectrometer), respectively. The neural network model is implemented and trained using the machine learning framework \textsf{TensorFlow} \cite{Tensorflow2016} version 2.4.3. For training, we adopt the weight initialization in \cite{He2015} and the Adam optimizer \cite{Kingma2014}. 80\% of the data are used for training with a minibatch size of 32, and the rest are used for testing. 

\begin{figure}[h]
	\includegraphics[width=0.5\textwidth]{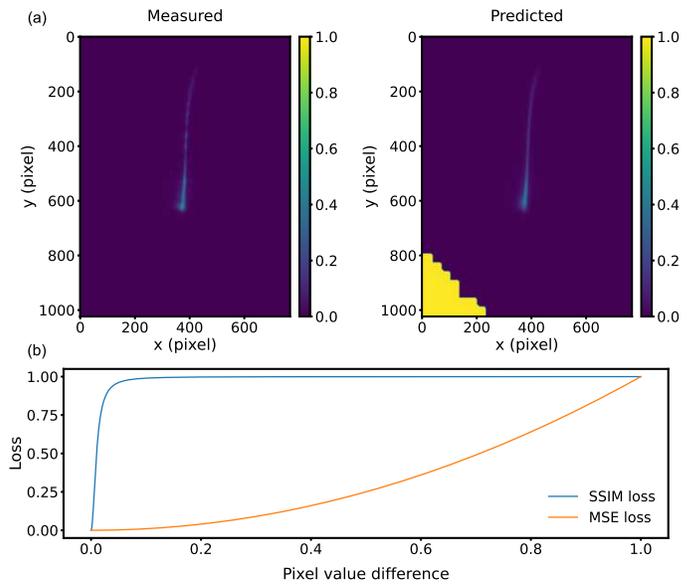}%
	\caption{\label{fig:loss_convergence} (a) Example of a prediction from a model trained using only the multiscale SSIM loss for the LPS image decoder. The pixel values near the lower-left corner of the predicted image are trapped in values around 1. This phenomenon occurs in all the LPS images. (b) Multiscale SSIM and MSE losses calculated between two unary images as a function of the pixel value difference. The pixel value of the first image is 0 while the pixel value of the second image ranges from 0 to 1.}
\end{figure}

The loss function $L_{total}$ for training is given by
%
%
\begin{equation}
L_{total} = L_{LPS} + wL_{spectrum},
\end{equation}
where $L_{LPS}$ and $L_{spectrum}$ are the loss functions for the LPS image decoder and the spectrum decoder, respectively, and $w$ is the weight which balances the influences of the two decoders. The model is trained with a learning rate of $3 \times 10^{-4}$ for 400 epochs, and then a learning rate of $1 \times 10^{-4}$ for another 400 epochs. During the first 400 epochs, both $L_{LPS}$ and $L_{spectrum}$ are the mean squared error (MSE) and $w$ is set to 1. During the second 400 epochs, $L_{LPS}$ is changed to the multiscale structural similarity index measure (SSIM)  \cite{Wang2004} with hyperparameters defined in \cite{Zhu2021} and $w$ is set to 100 empirically in order to make the losses of the two decoders at the same order of magnitude at the end of the training. There are two advantages for using different loss functions for the LPS image decoder in different training phases. First, the multiscale SSIM loss is much more computationally expensive than the MSE loss. In our implementation, the training time for a single batch reduces by about 30\% when the MSE loss is used. Therefore, the multiscale SSIM loss is only used to fine-tune the model in order to learn high-frequency features in the LPS images \cite{Zhu2021}. Second, it is found that sometimes the model does not learn correctly near a corner or an edge for all the images when only the multiscale SSIM loss is employed, as shown in Fig.~\ref{fig:loss_convergence} (a). This can be explained by the gradients of both the loss functions, as shown in Fig.~\ref{fig:loss_convergence} (b). The pixel values near the lower-left corner of the predicted LPS image shown in Fig.~\ref{fig:loss_convergence} (a) result in an extremely small gradient of the multiscale SSIM loss on the plateau, which prevents them from converging towards 0. However, how those pixels are trapped on the plateau is not clear.

The performance of our model is reported separately for the LPS image decoder and the spectrum decoder using the single-scale SSIM and MSE as metrics, respectively, over the test dataset.

\begin{figure*}[ht!]
	\includegraphics[width=1.0\textwidth]{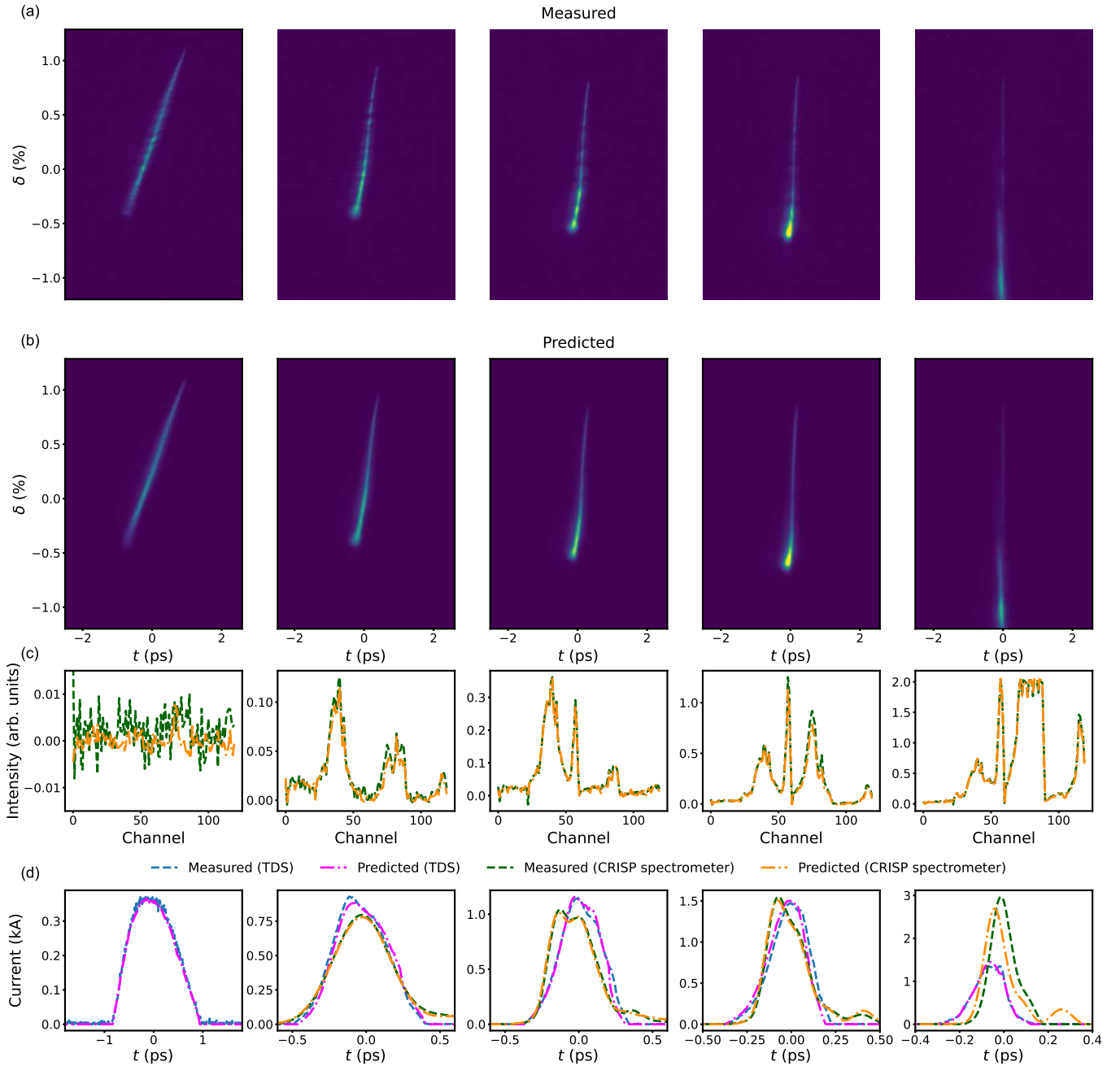}
	\caption{\label{fig:predictions_wp1} Five typical results from the test data of a dataset of WP1. The compression strength increases monotonically from left to right. (a) Measured LPS images. $\delta$ denotes the fractional energy deviation. (b) Predicted LPS images. (c) Comparisons between the measured and predicted spectra from the low-frequency grating set of the CRISP spectrometer. (d) Comparisons of the current profiles calculated from the measured and predicted LPS images as well as reconstructed from the measured and predicted spectra.}
\end{figure*}

\section{Results}

We record multiple datasets during two different beam times with different machine setups (working points). The beam energy is $\sim$1124 MeV for the first working point (WP1) and $\sim$1027 MeV for the second one (WP2). The bunch charge is $\sim$400 pC for both the working points. As mentioned previously, the beam parameters are not optimized for the LPS measurement due to the parallel FLASH2 user experiment. Consequently, it is found that the LPSs depend on the streaking direction of the TDS in a non-trivial way for WP2 due to transverse-longitudinal correlations \cite{Emma2000, Schmidt2020}, which makes it difficult to compare the current profiles measured by the TDS and the CRISP spectrometer. Nonetheless, this does not affect the conclusion of this study because the goal is to achieve an excellent agreement between the prediction and the measurement. Actually, the agreement between the current profiles measured by the TDS and the CRISP spectrometer can be verified using the dataset of WP1.

\begin{figure}[h]
	\includegraphics[width=0.5\textwidth]{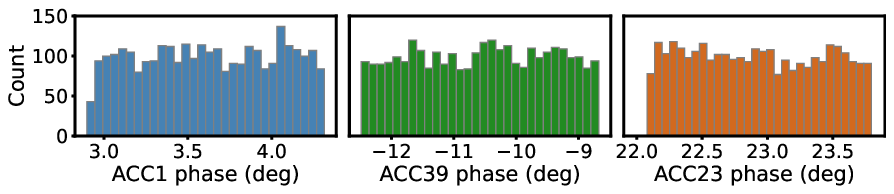}%
	\caption{\label{fig:input_histograms_wp1} Histograms of the scanned parameters for the dataset shown in Fig.~\ref{fig:predictions_wp1}. The total number of data points is about 3000.}
\end{figure}

\subsection{\label{prediction} Prediction}

Five typical prediction results from a dataset of WP1 are shown in Fig.~\ref{fig:predictions_wp1}. The electron bunches have significantly different peak currents. The histograms of the scanned parameters are shown in Fig.~\ref{fig:input_histograms_wp1}. The performance of the LPS image decoder is $0.9877 \pm 0.00227$ and the performance of the spectrum decoder is $7.9 \times 10^{-5} \pm 4.6 \times 10^{-4}$. As shown in Fig.~\ref{fig:predictions_wp1} (a) and (b), the predicted LPS images all agree with the measured ones excellently. There are also excellent agreements between the measured and predicted spectra, as shown in Fig.~\ref{fig:predictions_wp1} (c).

The current profiles calculated from the measured and predicted LPS images as well as reconstructed from the measured and predicted spectra using the combination of analytical (Kramers-Kronig) and iterative phase retrieval methods \cite{Schmidt2020} are shown in Fig.~\ref{fig:predictions_wp1} (d). The spectra in the left-most case (weakly compressed) contain only noise. Therefore, no current profiles can be reconstructed from them. In the right-most case, due to considerable energy loss induced by the CSR effect, there are unknown portion of electrons off the screen. Therefore, the peak current calculated from the LPS image is much smaller than that reconstructed from the spectrum. For the other three cases, the current profiles calculated from the LPS images and reconstructed from the spectra agree reasonably well.

It is noticed that there are conspicuous density modulations in the measured LPSs while the predicted LPSs are rather smooth. The density modulation is indeed induced by the microbunching instability which is seeded by the shot noise of an electron bunch \cite{Huang2002, Saldin2002, Ratner2015, Qiang2017}: an initial small density modulation inside the electron bunch can result in sufficient energy modulation due to the longitudinal space charge force, which in turn causes larger density modulation in a magnetic chicane bunch compressor. Because of the poor longitudinal resolution of the TDS measurement, only very weak density modulation can be observed in the current profile calculated from the measured LPS of the longest bunch in Fig.~\ref{fig:predictions_wp1}. The reconstructed current profiles from the spectra also do not show any evidence of the density modulation due to the lack of high-frequency components in the spectrum. Because such density modulation is not deterministic, it cannot be predicted by the neural network model using only RF phases and amplitudes as input. However, it is worth mentioning that the non-deterministic density modulation does not degrade the performance of the model. It appears that the model tends to predict the averaged LPS in which the density modulation is smoothed out.  

\begin{figure}[h]
	\includegraphics[width=0.5\textwidth]{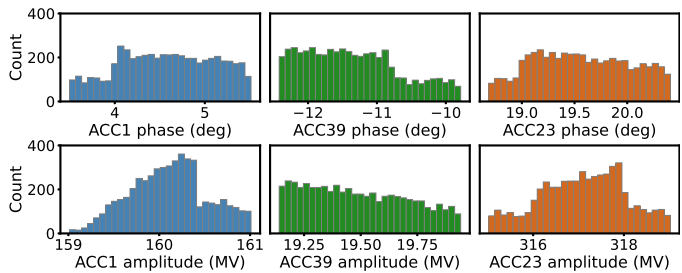}%
	\caption{\label{fig:input_histograms_wp2} Histograms of the scanned parameters for a dataset of WP2. The total number of data points is about 5200. The input data are not uniformly distributed because of two reasons. First, the data are combined from two datasets taken successively. The parameter ranges of the second dataset are smaller than the first one. Second, more than 40\% of the data are dropped, as discussed in Sec \ref{data_wrangling}. }
\end{figure}

\begin{figure}[h]
	\includegraphics[width=0.5\textwidth]{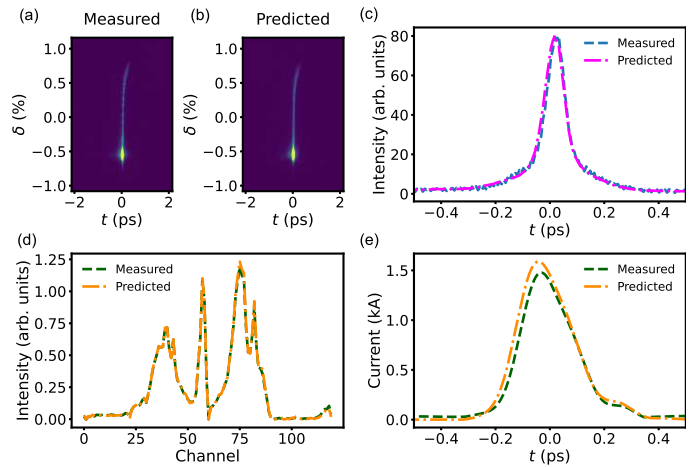}%
	\caption{\label{fig:predictions_wp2} An example result for the scan shown in Fig.~\ref{fig:input_histograms_wp2}. (a) Measured LPS image. (b) Predicted LPS image. (c) Comparison of the current profiles calculated from the measured and predicted LPS images. Because the LPS depends on the streaking direction of the TDS in a non-trivial way due to transverse-longitudinal correlations, the current profile here is simply the projection of the LPS image in the vertical direction. (d) Comparison of the measured and predicted spectra from the low-frequency grating set of the CRISP spectrometer. (e) Comparison of the current profiles reconstructed from the measured and predicted spectra.}
\end{figure}

In order to demonstrate the scalability of the model, six parameters are scanned for a dataset of WP2. The histograms of the scanned parameters are shown in Fig.~\ref{fig:input_histograms_wp2}. In total, more than 9,000 data points are taken. However, only about 5200 data points survive the data wrangling. Nonetheless, the performance of the LPS image decoder is $0.9801 \pm 0.00978$ and the performance of the spectrum decoder is $3.9 \times 10^{-4} \pm 3.2 \times 10^{-3}$. An example prediction is shown in Fig.~\ref{fig:predictions_wp2}.


\subsection{\label{mixed_diagnostics} Mixed diagnostics}

As mentioned previously, it is essential to combine the spectra from both grating sets of the CRISP spectrometer in order to achieve an accurate reconstruction of the current profile. At European XFEL, the high beam energy and short bunch length make it possible to extrapolate the result measured by the high-frequency grating set to the low-frequency regime \cite{Lockmann2021}. However, this is not feasible for the typical electron bunches at FLASH, especially when the electron bunches have a broad range of bunch lengths. Although another decoder could be trained to predict the spectrum measured by the high-frequency grating set, there are two concerns with this approach. Firstly, the spectra from the high-frequency grating set generally contain microbunching information, which fluctuates from bunch to bunch and train to train. Secondly, this will double the data collection time, which could be unaffordable for user facilities with very limited beam time for non-user experiments.

\begin{figure}[h]
	\includegraphics[width=0.5\textwidth]{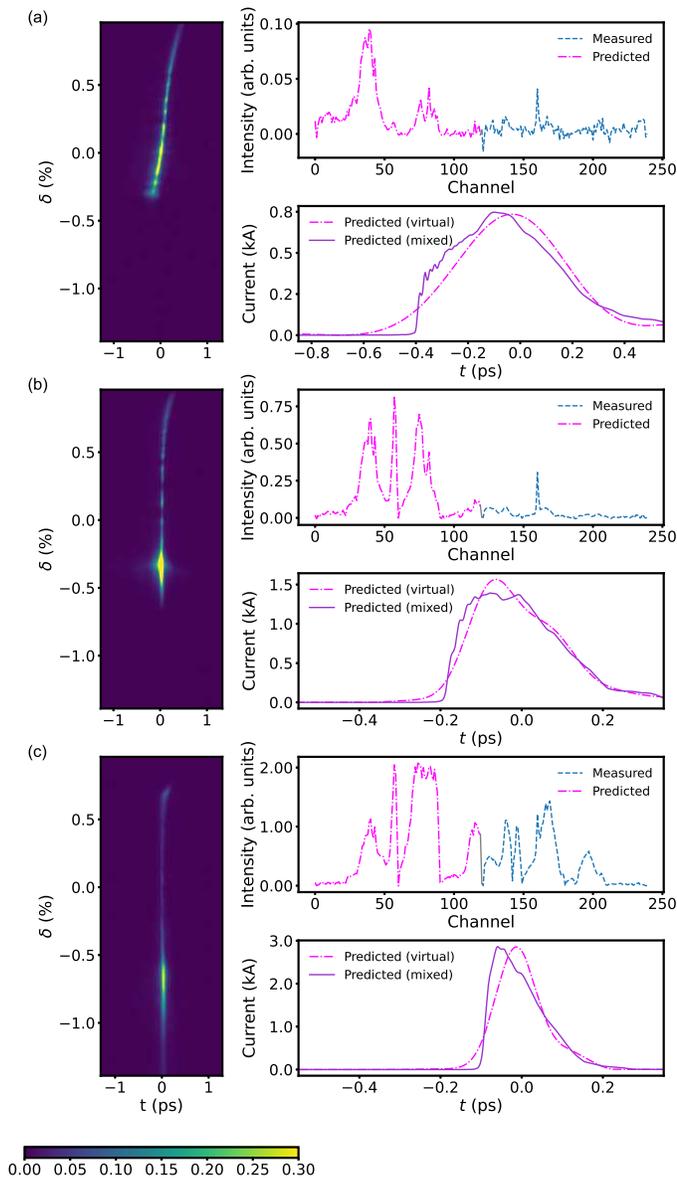}%
	\caption{\label{fig:predictions_mixed} Three typical results for mixed diagnostics with the CRISP spectrometer. The measured LPS image and the mixed spectrum are shown for each result. The current profiles reconstructed from the mixed spectra are compared with the current profiles reconstructed from only the predicted (virtual) spectra.}
\end{figure}

In order to improve the reconstructed current profile for real-time measurement, a mixed diagnostics method is proposed. The predicted spectrum for the low-frequency grating set can be combined with the measured spectrum from the high-frequency grating set so as to achieve the full-range spectrum in real time. To this end, we recorded two datasets using the high-frequency and low-frequency grating sets, respectively. Each dataset contains about 3000 data points. Only three control knobs (the phases of ACC1, ACC39 and ACC23) are scanned within the same ranges. A neural network model is trained using the dataset recorded with the low-frequency grating set. The performance of the LPS image decoder is $0.9854 \pm 0.00394$ and the performance of the spectrum decoder is $1.8 \times 10^{-4} \pm 1.3 \times 10^{-3}$. The trained model is also tested against all the LPS images in the dataset recorded with the high-frequency grating set and the performance of the LPS image decoder is as high as $0.9814 \pm 0.00525$. It indicates that the machine drift during data recording is negligible. Three typical prediction results with significantly different peak currents are shown in Fig.~\ref{fig:predictions_mixed}. They all show that it is essential to have the high-frequency components to reveal the exact shape and structures of the current profile although the low-frequency components are good enough to estimate the bunch length and the peak current.

\section{Conclusion}

In summary, we have experimentally demonstrated highly accurate megapixel LPS images and CTR spectra predictions for electron bunches in a bunch train concurrently at the end of the FLASH linac. Up to six major control knobs for LPS manipulation are scanned in order to collect data for electron bunches with broad ranges of LPS shapes and peak currents. The model is capable of providing heterogeneous LPS information and ensures reliable online diagnostics because a single type of diagnostic cannot cover electron bunches with a broad range of parameters. LPS images measured in the time domain are essential for electron bunches longer than a few hundreds of fs while CTR spectra provide more accurate current profiles for strongly compressed electron bunches.

A mixed diagnostics method is proposed to significantly improve the online current profile measurement using the CRISP spectrometer. The predicted CTR spectrum for the low-frequency grating set can be combined with the spectrum measured by the high-frequency grating set, which enables reconstruction of the current profile with a much higher accuracy in real time. Although the current profile reconstructed from the spectrum measured by the low-frequency grating set is good enough to estimate the bunch lengths and peak currents for typical electron bunches at FLASH, the spectrum measured by the high-frequency grating set is indispensable to reveal the exact shape and structures of the current profile. 

\begin{acknowledgments}
We would like to thank V. Ayvazyan, J. Branlard and S. Pfeiffer for advices on RF phase and amplitude scan, M. Hoffmann, J. Kral, J. Roensch-Schulenburg, C. Schmidt and M. Vogt for assistance as well as useful discussions during the data collection campaigns. We are grateful to the FLASH team for support and to B. Steffen for his thoughtful comments. 
\end{acknowledgments}

\bibliography{2021flash_vd}
\end{document}